# GIS-Based Estimation of Seasonal Solar Energy Potential for Parking Lots and Roads

Vishnu Mahesh Vivek Nanda
Center for Geospatial Analytics
North Carolina State University
Raleigh, NC, USA
vvivekn@ncsu.edu

Laura Tateosian
Center for Geospatial Analytics
North Carolina State University
Raleigh, NC, USA
lgtateos@ncsu.edu

Perver Baran
Center for Geospatial Analytics
North Carolina State University
Raleigh, NC, USA
kperver@ncsu.edu

*Abstract*— The amount of sun cast on roads and parking lots determines the charging opportunities for solar vehicles and impacts the efficiency of conventional vehicles. Estimates of solar energy potential on urban surfaces to assess parking and driving conditions need to account for the shadows cast by surrounding trees and buildings. However, though existing GIS tools can calculate solar potential on surfaces that have buildings and trees, these tools do not estimate the conditions beneath trees and do not consider the seasonal changes in deciduous trees. We introduce a new approach to address these factors using pixel substitution and a light penetration factor. In this paper, we describe how to integrate these techniques into a workflow for computing solar potential estimates for parking and driving conditions. We demonstrate the methodology in an urban setting in North Carolina that includes a mixture of urban structures and trees. We provide code samples so that this workflow is easily repeatable. The solar maps produced with our method are a useful resource for planning solar vehicle parking and routing, and identifying shaded conditions for conventional vehicles.

## I. Introduction

The declining price of Photovoltaic (PV) cells is driving the automobile industry to look towards solar energy as a potential alternative to fossil fuels. This is paving the way to the inception of Solar Electric Vehicles (SEVs) [1]. These vehicles harvest solar energy via PV cells affixed to their bodies and store it as electricity in batteries. A commercial SEV fueled by electricity produced with its solar panels is expected to be released next year [1]. These cars need direct sunlight to maximize charging. Large buildings and trees in urban landscapes block solar radiation, reducing the potential for charging parked or traveling SEVs. Some parking lots and roads are more suitable than others for SEVs, depending on the form of nearby buildings and trees. Identifying solar energy potential of parking lots and roads within a city will help decision-makers to plan for optimal SEV parking and movement within the city. This work presents an approach to model solar potential for this particular application as it considers buildings, tree structures, and seasonal changes for trees in an urban context.

Existing research has extensively explored the estimation of solar energy potential over building rooftops using GIS or GIS based tools [2]–[5]. Other studies have explored the solar energy potential of a topography at various scales, i.e., from a very small study area spanning a few square kilometers to an entire nation [6]–[9]. These studies have used either station-based (field) measurements or GIS-based tools, such as the solar irradiation functionality in GRASS GIS [6]. Studies using station-based measures assume a flat terrain with no consideration for shadows cast by the terrain or objects, such as buildings or trees. The studies that use GIS-based tools consider only the shadows cast by the terrain and buildings and ignore features, such as vegetation on the surface. Neither the field-based measurements nor the GIS-based tools are designed to model trees (or their shadows) over a topography. Furthermore, previous studies have not considered the seasonal changes in deciduous tree shadows. Therefore, a system is needed to model tree shadows over roads and parking lots to move toward better estimates of solar energy potential.

This paper presents a new approach that takes into consideration the surface vegetation, namely trees, when estimating solar energy potential of city parking lots and roads. The information gleaned from this process can help identify desirable parking lots and roads for SEVs and find shadier parking for parking conventional vehicles, information that can support decisions about allocating SEV parking spaces. This approach to estimating solar energy potential of city parking lots and roads uses one technique to estimate the solar energy potential of parking lots and roads for summer and another technique for winter. To demonstrate our proposed methodology, we apply our analyses over a 9 km$^2$ area of downtown Raleigh, NC, USA. This area contains a mixture of urban structures and vegetation, including both evergreen and deciduous trees.

The remainder of the paper is organized as follows. Section II describes the data and software required for our approach. Section III presents the proposed methodology in detail. In Section IV we apply this approach to downtown Raleigh data. Finally, Section V discusses the implications of our approach and the scope for future work.

## II. Data and Tools

The primary data source for this approach is LiDAR data. LiDAR data is collected with an airborne laser scanner system that generates a cloud of points where each point has its three-dimensional coordinates. Along with the coordinates, each point in a LiDAR dataset is usually associated with a few additional attributes, such as classification and return number. The 'classification' attempts to identify the type of object at that point, such as vegetation, building, water, and so forth. The 'return number' represents the number of the reflection out of multiple reflections of a laser pulse from which the point was recorded. The resolution and format of LiDAR data varies depending on the equipment used to collect it. A resolution of 2 points/m2 can be used for this type of solar radiation analysis, although higher resolution data will produce more precise results. Additional data is needed to designate the areas of interest for solar radiation. Our work is focusing on parking lots and roads in an urban environment, so it also requires data delineating key urban features (parking lots, roads, and buildings) in the format of shapefiles (a compound geographic file format that stores the features, such as points, lines, or polygons, as vector data). In addition, two types of imagery are required. One is needed for differentiating between deciduous and evergreen trees. This is a specialized geographic imagery dataset, orthorectified three-band multispectral imagery captured with an airborne sensor. The other imagery is needed for estimating light penetration through trees. These are photographs of tree crowns in the study area taken from ground level.

The primary software needed for this process is the free and open source software Geographic Information Software (GIS) tool, GRASS GIS (tested on Release 7.6) [10]. Specifically, it harnesses a solar irradiance and irradiation GRASS GIS module named r.sun [6], which is capable of estimating the solar energy potential over each pixel of an elevation model. LAStools[1] may be needed to support LiDAR data preprocessing. ForestCrowns, a USGS software tool, can be used to measure the light penetration of deciduous trees based on the canopy photographs [11].

## III. METHODOLOGY

The workflow for the proposed approach for estimating solar energy potential of city parking lots and roads is shown in Fig. 1. Given a LiDAR point cloud, a model of the terrain surface can be generated. The initial steps are to compute a digital surface model (DSM) and a digital elevation model (DEM) to be used as input into the GRASS GIS irradiation module. A DSM represents the features on the earth's surface as a raster image where each pixel value encodes the height of the object at that location. A DEM is similar to DSM except that it shows the height of the ground surface rather than any objects over it. DSM and DEM can be computed from 2 point/m² LiDAR data at a spatial resolution of 0.5 m. The DSM can be computed by assigning each pixel with the maximum height of the first returns of the LiDAR points within the pixel. The DEM can be computed using the minimum among the 'Ground,' 'Road surface,' and 'Bridge deck' classification categories of the LiDAR data. These elevation models can then be used as input to r.sun to estimate the solar energy potential over each pixel of the models. This tool is capable of estimating solar energy potential over individual time periods including yearly, monthly and daily. Even though this tool can also simulate cloudy conditions, we only focus on clear sky conditions. In this procedure, the 'daily' mode is used to estimate solar energy potential of parking lots and roads in two phases: leaf-on season and leaf-off season, referring to the seasonal change that occurs in deciduous vegetation.

### A. Leaf-on Season

The solar energy potential for a study area in the leaf-on season can be estimated by calculating the solar irradiance for a representative summer day using the GRASS GIS tool, r.sun, and pixel substitution. The r.sun tool can estimate the solar energy potential in metric units, watt-hour per square meter per day (Wh/m2/day) for any day of the year. The output is a raster map where each pixel stores the energy potential at that location [6]. The tool is designed to estimate the direct solar radiation, diffuse solar radiation, and ground reflected solar radiation falling over a unit area on any location on the earth. The solar energy potential over a unit area on the surface of the earth is the sum of these three components. The date of the summer solstice can be used for a conservative estimate of the leaf-on season conditions because vegetation greenery peaks at this time of the year.

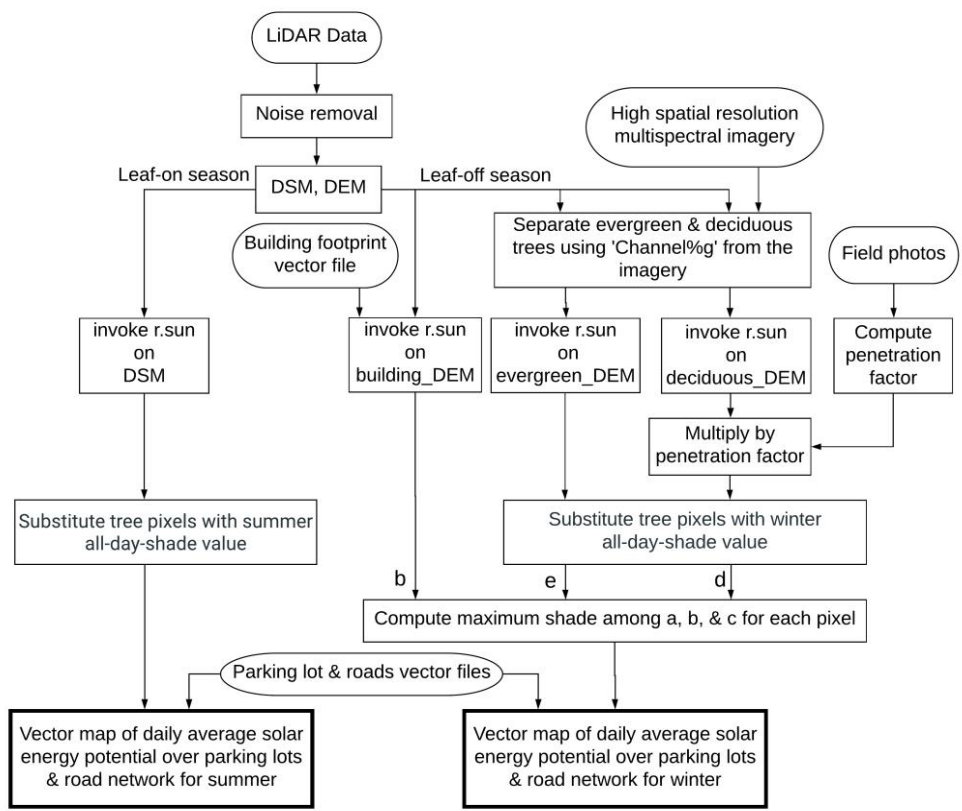

Fig. 1. Workflow: The process requires the following data as input: LiDAR point cloud, building footprints, vector files of parking lots and roads, high spatial resolution multispectral imagery and field photographs of deciduous trees (shown in ovals), and produces vector maps for the leaf-on and leaf-off seasons as output (in the bold boxes).

---

[1] https://rapidlasso.com/lastools/

To obtain the summer solstice solar energy potential over a study area, invoke r.sun with the study area DSM and this date in 'daily' mode. For our application, we are interested in the irradiation on the ground. A limitation of the output from this module is that wherever there are trees, the module produces the irradiance values that the top of the trees would receive, instead of the ground-level solar potential, which would be largely or entirely in the shade. To correct for this, the pixels directly below the trees need to be identified and the value can be modified to reflect day-long shade coverage (the most conservative estimate). These pixels can be identified by finding the tree crowns, because, in a 2D raster, the tree crown pixels correspond to the pixels directly beneath the crown.

To identify tree crowns, we assume that pixels that are above ground-level represent either vegetation or buildings, so that if we can rule out buildings, we are only left with differentiating between small vegetation (shrubs) and large vegetation (trees). We need to separate the latter two since vehicles are usually not able to park beneath small vegetation. We classify small vegetation as 2.5 m or less. Note that we are assuming that if you identify buildings and anything less than 2.5 m from the ground, the remaining above ground points represent the location of trees. This assumption is not valid for all applications because there are objects taller than 2.5 m that are neither trees nor buildings. However, since we are only focusing on parking lots, roads, and their immediate vicinity, it is a reasonable assumption that most objects taller than 2.5 m are trees or buildings.

Given these assumptions, locating a tree crown pixel can be performed using the building footprints of building data and a height range in the LiDAR data (to locate large vegetation). Though LiDAR data may contain a classification value for each pixel to identify the type of object at that point (i.e., ground, building, small, medium, or large vegetation), these values may have some building points classified as vegetation or the converse. Using building footprints that are more accurately surveyed and a LiDAR height range is a more reliable method for identifying tree crowns. To remove the buildings consideration, we first need to create a raster that encode the building locations. This can be created by intersecting the building footprint vector data with the DEM to select the DEM pixels that are within a building footprint. The corresponding DSM pixel values for these pixels can then be substituted into the DEM to obtain a new elevation model. We will refer to this raster as building_DEM. This elevation model encodes the elevation of the ground surface except within the footprints of buildings, where the elevation will be that of the building top. Using this DEM, a binary raster can be created to mark tree locations as follows:

$$tree\text{-}mask = \begin{cases} 1, \text{ if } (DSM - (building\_DEM)) > 2.5 \text{ m} \\ 0, \text{ otherwise} \end{cases}$$

Now the pixels under trees can be updated by substituting these pixels with an all-day shade value. First, the value, v, of a pixel that is known to be under shade throughout the day can be manually identified from inspection of the high-resolution imagery. Then, since tree-mask pixels set to 1 correspond to pixels beneath the crown, the values of these pixels in the r.sun output can be set to v. The resulting raster represents the modified estimate of the solar energy potential for the leaf-on season. It should be noted that in an output raster from r.sun, the value of a pixel that was under shade all throughout the day and the minimum value within the output will be the same.

*B. Leaf-off Season*

During the winter months, when deciduous trees shed their leaves, the bare branches of these trees allow light to penetrate to the ground much more than in leaf-on season. To get an estimate of the solar energy potential for parking lots and roads near these trees that takes into consideration this seasonal change, the percentage of sunlight penetrating through the deciduous tree structures needs to be calculated. To do this, first, we need to be able to differentiate between deciduous trees and evergreen trees, as most evergreens do not lose their leaves. Second, we can estimate the amount of uncounted solar potential available beneath the deciduous trees during leaf-off season and update our raster in the deciduous tree locations.

To determine the locations of deciduous and evergreen trees, LiDAR data and high spatial resolution three-band orthorectified aerial imagery captured in winter can be used. The winter imagery has to be captured during the winter season so that deciduous trees in the study area have shed their leaves. An imagery index, Channel%, can be calculated from this imagery to separate evergreen from deciduous trees [12]. The Channel% of the Green band within the imagery is calculated as,

$$Channel\%g = \text{Green}/(\text{Blue} + \text{Green} + \text{Red})$$

where Green, Blue, and Red are the three color bands in the imagery. The *Channel%g* values of pixels that contains evergreen trees will have higher *Channel%g* value than those that contain deciduous trees. The threshold value, t, that separates them can be determined manually through visual inspection. Next, using the *tree-mask* raster, calculated in Section III A, the *Channel%g* values of the tree pixels can be extracted. The tree pixel values that exceed t can be considered evergreen tree pixels (while the remainder are deciduous).

Once the evergreen trees and deciduous tree locations are identified, the effect of shadows cast by them needs to be computed individually. To accomplish this, two separate rasters, *deciduous_DEM* and *evergreen_DEM*, that represent the height of deciduous trees and evergreen trees respectively within the DEM can be created in a process analogous to *building_DEM*. The *evergreen_DEM* is generated by locating the evergreen pixels in the DSM and inserting these values in the evergreen positions in the DEM, so that *evergreen_DEM* encodes the ground surface, except for locations evergreens are detected. The same procedure applied to deciduous pixels creates *deciduous_DEM*. The r.sun tool can then be invoked separately on these to obtain their individual shading.

*1) Solar penetration through tree crowns:* The output of r.sun on the *deciduous_DEM* does not account for the penetration of light through the crowns of deciduous trees which have shed their leaves. A USGS tool, named ForestCrowns, can be used to address this[11]. This tool estimates canopy cover of trees based on input photographs of tree canopies taken from the ground. An example is shown in Fig. 2. For best results, Winn et al. (2016) recommend capturing input photography via a tripod-mounted vertical facing camera, standing where there are no low-hanging branches. Photos should be taken when the sky is clear and the sun is not directly overhead. This software calculates the ratio of sky pixels to the total number of pixels within a region of interest, as a measure of canopy cover. In this application, this software can be used to determine the transparency of the crown of a deciduous tree by using photographs of deciduous trees in the study area as inputs. The light penetration factor through the deciduous trees in the study area can be calculated as the mean value of the ratios obtained from a sample of photographs.

To integrate the results from the ForestCrown process, the light penetration factor needs to be applied to the *r.sun* raster output for *deciduous_DEM*. However, time under the sunlight must be taken into account. Some areas will be under the shade of a deciduous tree for longer than others and these areas should not be modified for light penetration by the same magnitude. A pixel's time under sunlight is proportional to the value of that pixel in the *r.sun* output. The difference between an all-day sunlit pixel (maximum pixel value) and the pixel value at any location is a measure of the duration for which the pixel was under sunlight. Therefore, the pixels within the *r.sun* output of *deciduous_DEM* with a value less than the overall maximum pixel value should be recalculated. It should be noted that any two pixels within *r.sun* output under the same sunlit conditions may show fractionally different values and this difference is within 0.5% of the maximum value. Thus, the pixels within the *r.sun* output of *deciduous_DEM* with a value less than 99.5% of the overall maximum pixel value should be recalculated using the formula

$$\text{new\_pixel\_value} = \text{current\_value} + \text{light\_penetration\_factor} * (\text{max} - \text{current\_value})$$

where max is the maximum pixel value of the r.sun output of deciduous_DEM.

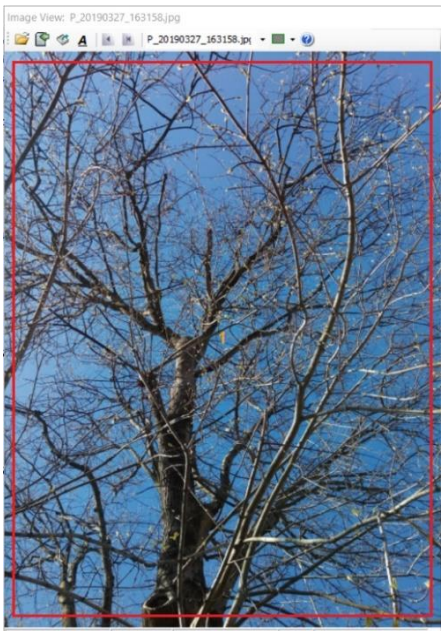

Fig. 2. A tree photograph loaded into the ForestCrowns interface showing the region (in red) to be analyzed.

*2) Solar energy potential directly beneath trees:* The *r.sun* outputs of *evergreen_DEM* and light penetration adjusted *deciduous_DEM* are now available. Like the case of leaf-on trees, the pixel values directly beneath trees have to be adjusted. Unlike leaf-on season, here we have two types of trees: the evergreen trees (which can be treated similar to leaf-on case) and deciduous trees. In the case of evergreen trees, we first find the pixels that are beneath the evergreen trees using the evergreen *tree-mask* that we created earlier. Next, we replace those pixels with the value of a pixel that was under shade the whole day. In the case of deciduous trees, we first find out the pixels that are beneath the deciduous trees using the deciduous *tree-mask* that we created earlier. Next, we replace the pixel values in the modified *r.sun* output of *deciduous_DEM* with new values that correspond to deciduous trees by using the formula

$$\text{new\_pixel\_value} = \text{min} + \text{light\_penetration\_factor} * (\text{max} - \text{min})$$

where min is the minimum and max is the maximum value of *r.sun* output of *deciduous_DEM*.

There are some scenarios where a tree's shade will be engulfed by a building shade or the shade of a deciduous tree may overlap with the shade of an evergreen tree. In those cases, it is obvious that the resultant shade at a particular point will be cast by one which is blocking the sunlight the maximum. When the blockage of sunlight falling on a pixel increases, the solar energy potential over the pixel decreases. So in our case, the resultant total solar energy potential output raster is obtained by assigning each pixel a value which is the minimum among the three modified *r.sun* outputs.

IV. APPLICATION

To demonstrate our approach, we apply it here to a study area in downtown Raleigh, NC, that has a mixture of urban structures and vegetation, including both evergreen and deciduous trees. We gathered LiDAR data, parking lot, road and building data, and imagery for the study area. For the LiDAR data, we used 2015 North Carolina Quality Level 2 (QL2) LiDAR data with a resolution of 2 points/m$^2$. Raleigh resides within the political boundaries of Wake County and for the parking lot and roads we used polygon vector data from Wake County's open data portal[3]. Building footprints vector file for the study area was extracted from Microsoft's US building footprints database[4]. The building footprints are two-dimensional. In other words, no height values are available within this dataset. For aerial imagery, we used six-inch spatial resolution, orthorectified, three-band multispectral imagery of the study area, which is freely available from the geographic data portal of North Carolina[5]. As a proof of concept, we captured five photographs of leaf-off deciduous tree crowns from different parking spots to represent deciduous trees in the study area.

First, we used LAStools to remove noise from the LiDAR data. For the Leaf-on season, we ran *r.sun* using DSM created from the LiDAR data for June 21st, the date of summer solstice, to obtain the solar energy potential over the study area (refer Fig 3.a). The solar energy potential associated with a pixel that was under shade all day long in our study area was found to be 1,216 Wh/m²/day. Hence, all the pixels of the output from *r.sun* that represented trees were substituted with the value of 1,216 Wh/m²/day to obtain the final solar energy potential estimate for the leaf-on season (refer Fig 3.b).

---

[2] https://sdd.nc.gov/
[3] http://data-wake.opendata.arcgis.com/
[4] https://github.com/microsoft/USBuildingFootprints
[5] https://www.nconemap.gov/

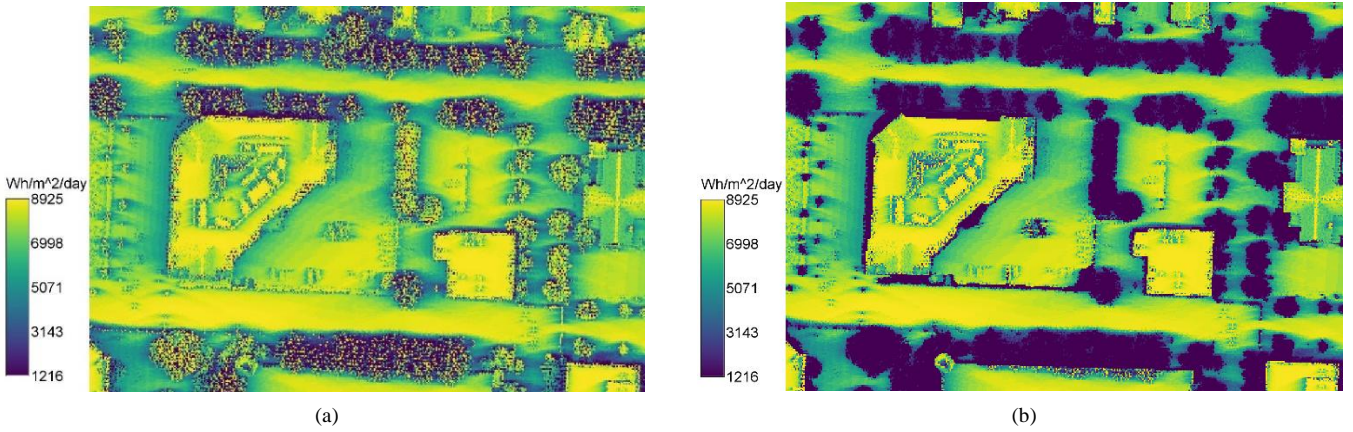

Fig. 3. Solar energy potential estimates for a Raleigh city block (r.sun calculated on the DSM): (a) before removing trees, and (b) after trees substituted with a 1,216 Wh/m²/day pixel value, the accumulated indirect irradiation of day-long summer shade in the study area.

For the leaf-off season, we first created the building_DEM using LiDAR data and the Microsoft building footprints layer. Then we computed the solar irradiance, *b*, of this raster by running r.sun with a mid-winter date, January 1. The maximum and minimum pixel values of the *r.sun* output were 3,105 Wh/m²/day and 678 Wh/m²/day, respectively. We then computed the *Channel%g* for the study area from the six-inch spatial resolution multiband imagery and found the threshold for evergreen trees to be 0.375. We then ran *r.sun* with *evergreen_DEM*, and *deciduous_DEM* as inputs. Following the same procedure as in the leaf-on season, the pixels directly beneath the trees in the output of *r.sun* with *evergreen_DEM* were substituted with the minimum pixel value of the same raster (678).

Since the output of *r.sun* with *deciduous_DEM* doesn't account for light penetration through deciduous trees, we used the deciduous tree crown field photographs as input in ForestCrowns software and calculated the transparency of the tree crowns (as shown in Fig 2). The average crown transparency of the five deciduous trees was approximately 2/3 of the analyzed photograph regions, so a value of 0.67 was used for the light penetration factor. This light penetration factor was applied to the *r.sun* output of *deciduous_DEM* as:

new_pixel_value = current_pixel_value + 0.67 * (3,105 – current_pixel_value)

The shade beneath the deciduous trees were accounted for in the *r.sun* output for *deciduous_DEM* as

new_pixel_value = 678 + 0.67 * (3,105 - 678)

The resulting raster *d* was then merged with the other two outputs, the rasters labeled *e* and *b* in Figure 1. While merging, it was made sure that the result has the minimum value among the corresponding pixels from these three rasters.

The solar energy potential over our study area is obtained as two rasters, one for the leaf-on season (June 21) and the other for the leaf-off season (January 1). Each pixel of these rasters represents the total solar energy for a day in Watt-hours/m²/day. Since we are only interested in the solar energy falling over parking lots and roads, the pixels that belong to parking lots and roads were extracted from these rasters using the parking lot and road network data. The average of the extracted values within each parking lot and road network segments is used as the solar energy potential of parking lots and road network in the study area. The output for parking lots is shown in Fig 4. With the added information about tree shade patterns, we are able to observe some interesting patterns. In the summer, the solar potential is much more variable than in the winter, perhaps due to leaf coverage. Some parking lots which had low solar energy potential in the summer jumped up in ranking during winter. Left center in Fig. 4a, you can see parking lots that were in the lowest leaf-on solar class (yellow) that are ranked much higher (middle class) in the leaf-off season, Fig. 4b. Note that, though the reduced leaf cover in the winter allows more of the available light to pass through the trees, the values for leaf-off season (Fig. 4b) are lower because the solar irradiation is lower in winter than during the summer (Fig. 4a).

## V. Discussion and conclusions

In this work, we have presented a new approach to solar energy estimation which incorporates the variations in the shadows cast by trees. This method identifies tree locations and distinguishes between deciduous and evergreen trees so that distinct treatments can be applied to these tree types. Pixel substitution was introduced to account for shade directly below both types of trees. For deciduous trees in leaf-off season, our workflow derives a light penetration factor from field photographs. We derived the penetration factor from field photographs.

We demonstrated our methodology in an urban setting and identified distinct differences in favorability of parking lots within the study area for solar vehicles and conventional vehicles. In this study area, the contrast in favorability is greater in leaf-on season than leaf-off season. These observations would not be as visible without the introduction of pixel substitution and light penetration factors into the calculations.

Our methodology is easily repeatable for other urban study areas where data is available. Sample code for the procedure that produced this output is available on Github[6]. The software tools used to complete the process are either free and open-source (GRASS GIS) or free (ForestCrowns). Further, we have provided open source sample code that can be customized for new study areas. The output rasters where each pixel represents the solar energy potential over it was extracted and later visualized using vector files of parking lots and road network of the study area. This will help the decision makers to make informed decisions for optimal parking and movement of SEVs in the area. Information such as this can be made available to applications that can inform SEV users to make optimal parking decisions and monitor estimated charge.

---

[6] https://github.com/eternalscholar/Solar_Potential_Project

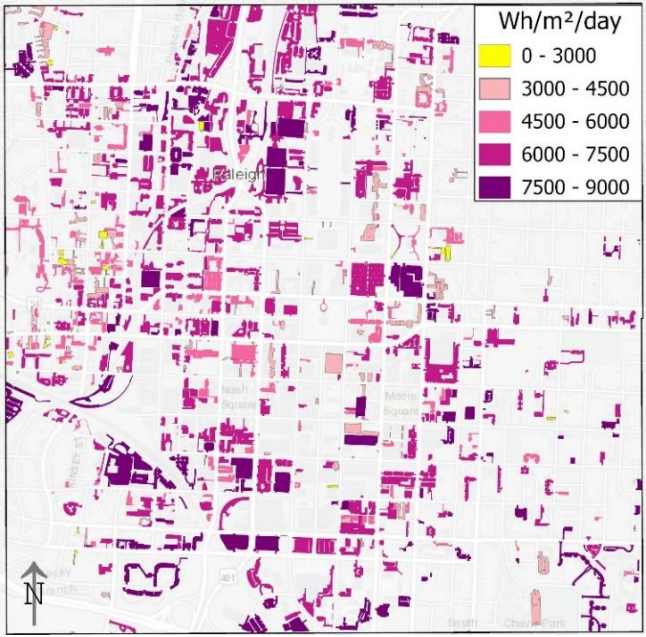 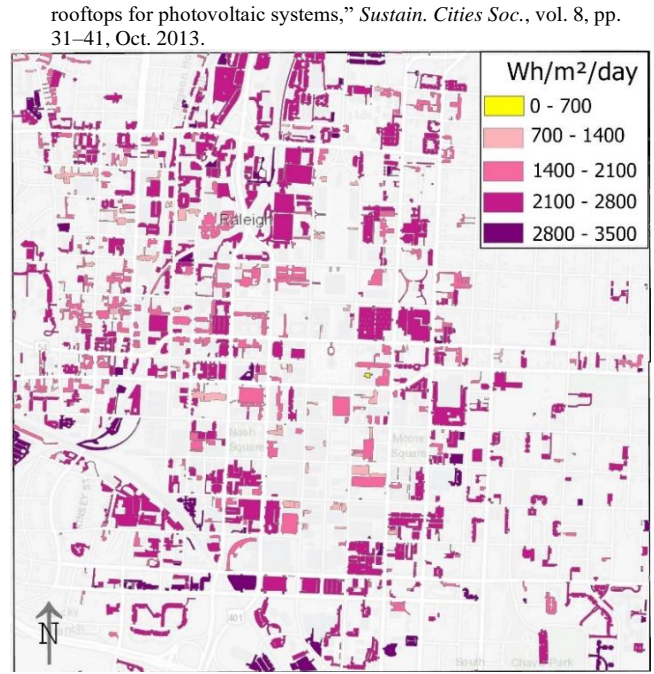

Fig. 4. Solar energy potential of parking lots within the study area (a) for leaf-on season and (b) leaf-off season.

Though we used existing data for buildings, roads, and parking lots to demonstrate our methodology, machine learning algorithms can be used to generate these datasets from aerial imagery. Object recognition also has implications for extending our methodology. Trees vary in size, shape, leaf area, leaf density, branch density, branch thickness. An exciting direction for future work includes harnessing deep learning to determine the species of trees in the study area so that we can introduce customized tree models into our workflow and derive multiple penetration factors.